\documentclass[twoside]{article}

\usepackage{amsmath}
\usepackage{amsfonts}
\usepackage{amssymb}
\usepackage{graphicx}
\usepackage{a4wide}
\usepackage{times}     

\def\date{February 28, 2004}

\newcommand{\ed}{\end{document}}
\newcounter{mycnt}[section]

\def\nn{\nonumber \\}
\def\Id{\text{I\!d}}
\def\antip{\textsf{S}}
\def\openk{\Bbbk}
\def\openR{\mathbb{R}}
\def\openC{\mathbb{C}}
\def\openN{\mathbb{N}}

\def\!{\kern -0.15ex}
\def\nn{\nonumber\\}
\begin{document}
\title{\Large\textsf{Products, coproducts\\[0.5ex] 
and singular value decomposition}}
\author{Bertfried Fauser$^{1)}$}
{{\renewcommand{\thefootnote}{\fnsymbol{footnote}}
\footnotetext{\kern-15.3pt AMS Subject Classification: 15A18; 16W30; 15A66.\par
\vskip 8pt\noindent\vbox{
\noindent\small
{\sc 1) Max Planck Institut f\"ur Mathematik, Inselstrasse 22-26, D-04103 
Leipzig, Germany, E-mail: \tt Bertfried.Fauser@uni-konstanz.de}
}}}
\maketitle
\begin{abstract}
Products and coproducts may be recognized as morphisms in a monoidal tensor 
category of vector spaces. To gain invariant data of these morphisms, we can
use singular value decomposition which attaches singular values, \textit{ie} 
generalized eigenvalues, to these maps. We show, for the case of Grassmann and 
Clifford products, that twist maps significantly alter these data reducing 
degeneracies. Since non group like coproducts give rise to non classical 
behavior of the algebra of functions, \textit{ie} make them noncommutative, 
we hope to be able to learn more about such geometries. Remarkably the coproduct
for positive singular values of eigenvectors in $A$ yields directly corresponding
eigenvectors in $A\otimes A$.

\end{abstract}
\tableofcontents

\section{Introduction}

It is well known that function algebras on group manifolds can be recast
in a Hopf algebraic setting. The famous Gelfand theorem tells us that
every commutative C-*-algebra is dual to the algebra of functions on some 
topological function space under point wise multiplication. Hence the geometric
data can be handled either in the algebraic or in the function theoretic setting.

Since noncommutative C-*-algebras occur naturally. It was an obvious question 
to ask, which type of geometries are related to the dualized function algebras. 
However, these function algebras have to be noncommutative. One idea behind
this mechanism is the following. Assume there is a point $x$ in a manifold $M$.
We try to find the value of the product of two functions $f,g : M \rightarrow \openC$
on $x$
\begin{align}
(f\ast g)(x) &= f(x)g(x)
\end{align}
using the \textit{point wise} multiplication of the function values. In other words, 
the product of two functions is dual to the coproduct on the points of the manifold. 
Symbolically
\begin{align}
(f\ast g)(x)&= (m(f\otimes g))(x) \,=\, (f\otimes g)(m^*(x))
\,=\, (f\otimes g)(x\otimes x) \,=\, f(x)g(x).
\end{align}
Here we had to assume that the coproduct $m^*$ is group like, \textit{ie} 
$m^*(x)=x\otimes x$, and that the evaluation map $\textsf{eval}(f\otimes x)=f(x)$ 
is generalized canonically as a crossed map to $\textsf{eval}_{(V^*)^{\otimes^2}
\otimes V^{\otimes^2}}((f\otimes g)\otimes (x\otimes y)) = f(x)g(y)$. While this
mechanism seems to be natural, since we have used it already in high school, it can
readily be generalized to the case where one demands that the coproduct is non-group
like. In \cite{chryssomalakos:1998a} one may look up a detailed description of that
point of view, and what may change in the underlying geometry. 

A second source of noncommutativity is related to twist maps and 'quantization'
\cite{fauser:1996c,fauser:2002c,hirshfeld:henselder:2002a}. Such twist maps can be 
subsumarized under the name of \textit{cliffordization}. An alternative name
would be comodule algebra map. We prefer the former in combinatorially intended settings.
The twisted product of two morphisms is given as
\begin{align}
f\ast_\chi g &= \sum_{(f)(g)} \chi(f_{(1)},g_{(1)}) \, f_{(2)}\ast g_{(2)}
\end{align}
where we have used Sweedler indices \cite{sweedler:1969a}. It is easily seen, that this
leads in general equally well to a non-commutative function algebra.

An easily tractable and universal model of such a deformation is the transition from the 
Grassmann Hopf algebra (or symmetric Hopf algebra) to the Clifford comodule algebra (or 
Weyl comodule algebra) as described in detail in \cite{rota:stein:1994a}. Hence it might 
be useful to skip all further complications and to investigate the product and coproduct
structure in such algebras. A natural way to study such deformations is using 
cohomological methods \cite{sweedler:1968a}. This led to amazing insights into the 
structure of quantum field theories \cite{brouder:fauser:frabetti:oeckl:2002a} and 
symmetric functions \cite{fauser:jarvis:2003a}. While this method produced even 
computational tools and is suited for super algebras etc, we want to take in this 
paper another route.

Any product in an algebra $A$ is a linear morphisms $m : A\otimes A \rightarrow A$. Seen 
in the category of modules, its just a module morphism from $B=A\otimes A$ to $A$. Hence,
assuming finite dimensionality for the sake of simplicity, and introducing bases, we 
get a rectangular representation matrix for a product morphisms, \textit{ie} the
multiplication table. Let $\{a_i\}$ be a basis of $A$ and $\{b_I\cong (a_i\otimes a_j)\}$
a basis of $B=A\otimes A$, we get
\begin{align}
m(a_i\otimes a_j) = m(b_I) = \sum_k m_I^k\, a_k = \sum_k m_{ij}^k a_k \, . 
\end{align}
The change of perspective of formally reducing a tensor of degree three to a 
tensor of degree two has to be payed for by dealing with rectangular matrices. 
The same holds true for coproducts, where we easily see, that products and coproducts
of a Hopf algebra obeys representations\footnote{%
We assume $A$ to be of dimension $n$, the dimension of $B$ is then $m=n\times n$.
However, our arguments run through without this speacialisation for arbitrary 
spaces $A$ and $B$.}
as $n\times m$ and $m\times n$ matrices, which allows to concatenate them.
Notationally, we will use lower case indices for elements in $A$ and upper case
indices for elements in $B=A\otimes A$. Let $\alpha$ be the isomorphism which
encodes and decodes the two index sets, hence
\begin{align}
\label{alpha}
\alpha(I) \,=\, [i,j] &&&\alpha^{-1}([i,j]) \,=\, I \, .
\end{align}
A matrix representation of $\alpha$ is a tensor of degree three 
$\alpha^I_{ij} \in B \otimes (A^*\otimes A^*)$.
Having done this, we can apply the techniques from ordinary linear algebra, \textit{ie}
singular value decomposition to characterize product and coproduct maps. We will see in
the course of this work, that this information is more subtle and detailed then the 
cohomological classification, and therefore opens up new theoretical insight. Moreover,
it is well known from singular value theory, that the large singular values characterize
a rectangular map reasonably well. Hence we might hope to expand products and coproducts
using only a few large singular values, dropping small ones without great loss of 
information. In this way, we hope to develop a method, which will allow to replace 
complicated product and coproduct structures in a coherent way, \textit{ie} maintaining
the Hopf algebra structure, by a much simpler and well adapted product coproduct pair.
Ultimately, we hope to get geometrical insights via this approach as well.

\section{Hopf algebra structure}

To make this paper more self contained and not to assume much knowledge about Hopf
algebras, we will provide the axioms here. Some of the formulas are needed for reference
issues also. References for Hopf algebra theory may be
\cite{sweedler:1969a,abe:1980a,kassel:1995a,majid:1995a}

Let $A$ be an associative, unital $\openk$-algebra. We denote the underlying 
$\openk$-module of $A$ by abuse of notation also with $A$. The product map is denoted 
$m : A\otimes_\openk A \rightarrow A$ and is a $\openk$-linear map of modules in the
monoidal category of $\openk$-modules $\textsf{mon}_\openk$. The unit is 
$\eta : \openk \rightarrow A$. The monoid forms a symmetric tensor category with respect
to the twist map $\textsf{sw} : A\otimes B \rightarrow B\otimes A$. Note that the switch
map has to be universal (natural), hence one has to impose a coherence law which in this
case is given by the braid equation. For our purpose important is the fact, that the
switch map is represented as a permutation matrix $\textsf{P}$ under the $\alpha$
isomorphism
\begin{align}
\alpha\circ\textsf{sw}(A\otimes A) &= \textsf{P}\circ \alpha(A\otimes A) 
\,=\, \textsf{P}\alpha(B) \, .
\end{align}

Let $C$ be an coassociative, counital $\openk$-coalgebra. we denote the underlying 
$\openk$-comodule of $C$ by abuse of notation also with $C$. The coproduct map is
denoted $\delta : C \rightarrow C\otimes C$ if group like and $\Delta : C \rightarrow
C\otimes C$ if not group like. The counit is $\epsilon : C \rightarrow \openk$. $\delta$ 
and $\Delta$ are morphisms in $\textsf{mon}_\openk$. We adopt the Brouder-Schmitt 
convention \cite{brouder:schmitt:2002a}, denoting the Sweedler indices of the coproduct 
of $\delta(x)=x_{[1]}\otimes x_{[2]}$ and $\Delta(x)=x_{(1)}\otimes x_{(2)}$ using
different bracings. 

An algebra $A$ (coalgebra $C$) is called augmented, if it has an (co)augmentation
morphisms, a counit $\epsilon : A \rightarrow \openk$ (an unit 
$\eta : \openk \rightarrow C$). An (co)augmented (co)algebra is called connected,
if the (co)augmentation as an (co)algebra map satisfies
\begin{align}
\epsilon\circ m \,=\, m_\openk\circ(\epsilon\otimes\epsilon)
&&&
\Delta\circ\eta \,=\, (\eta \otimes \eta)\circ \delta^\openk
\end{align}
It is known, that twists of connected (co)algebras lead in general to nonconnected 
(co)algebras, even if the twist is cohomologically trivial, \textit{ie} induced via
a 2-coboundary \cite{brouder:fauser:frabetti:oeckl:2002a,fauser:jarvis:2003a}.
Such algebras were coined 'interacting' in \cite{fauser:2002c}.

A bialgebra\footnote{%
We use the common letter $B$ for bialgebra, not to be confused with the $B$ 
intoduced above.} 
is a module $B$ carrying an algebra and a coalgebra structure, such that the
compatibility law
\begin{align}
\label{AXIOM1}
\Delta\circ m &= (m\otimes m)(\Id\otimes\textsf{sw}\otimes\Id)(\Delta\otimes\Delta)
\end{align}
holds. This states that $m$, ($\Delta$) is a coalgebra (algebra) homomorphism. This 
compatibility law allows actual computations, since it embodies the germ of Laplace
expansions, together with the dual Hopf algebra.

A Hopf algebra $H$ is a bialgebra where an antipode $\antip : H \rightarrow H$ exists
fulfilling
\begin{align}
\antip(x_{(1)})x_{(2)} &= \eta\circ\epsilon(x) 
\,=\, 
x_{(1)}\antip(x_{(2)}) \, .
\end{align}

It is possible to start with the convolution demanding the existence of an antipode. 
It was proved by Oziewicz, that any antipodal convolution has a crossing which fulfills
eqn. (\ref{AXIOM1}). However, the crossing needs not to be the switch and even not to be
be a braid. Such algebras were denoted \textit{Hopf gebras}, see \cite{fauser:2002c}
for details.

\section{Grassmann Hopf algebra and twists}

To simplify our discussion, we will study Grassmann Hopf algebras, which are
computationally manageable and provide nevertheless an archetypical example. Let $V$
be a finite dimensional vector space, the exterior powers of $V$ are denoted as 
$\lambda^r(V)=V^{\wedge^r}$, which is a graded space $\Lambda(V)=\sum \lambda^r(V) =
\sum V^{\wedge^r}$. The product is given by the exterior product $\wedge$ 
(wedge product) and the coproduct is given recursively by
\begin{align}
\Delta(v) &= v\otimes \Id + \Id \otimes v\hskip 2truecm \text{$v$ in $V$} \nn
\Delta(A\wedge B) &= \sum \pm A_{(1)}\wedge B_{(1)}\otimes A_{(2)}\wedge B_{(2)}
\,=\,Delta(A)Delta(B)
\end{align}
where the sign is given by the alternating character of the symmetric group yielding
the graded switch $\textsf{sw}(V^{\wedge^r} \otimes V^{\wedge^s}) = (-1)^{rs}
V^{\wedge^s}\otimes V^{\wedge^r}$ for the crossed products and extended by linearity. 

The antipode is given as $\antip(V^{\wedge^s})=(-1)^s\, V^{\wedge^s}$ and the counit 
is is given as $\epsilon(\Id)=1$, $\epsilon(V^{\wedge^r})=0$ for all $r>0$.

Let $\{e_i\}$ be a basis of $V$, a basis for elements of $V^{\wedge^r}$ is given by
$\{e_{i_1}\wedge e_{i_2}\wedge \ldots \wedge e_{i_r}\}$ where $i_1<i_2<\ldots< i_r$.

We can now introduce a new product, called cliffordization or circle product, using
a general bilinear form $B^\wedge$ on $V^\wedge\otimes V^\wedge$ induced from a 
bilinear form $B : V\otimes V \rightarrow \openk$ as
\begin{align}
\label{cliffordization}
x\circ y &= \sum_{(x),(y)}\,\pm\, B^\wedge (x_{(1)},y_{(1)})\, x_{(2)}\wedge y_{(2)}\, .
\end{align}
The bilinear form is evaluated by Laplace expansion
\begin{align}
\label{laplaceB}
B(\openk,V) \,=\, 0 &= B(V,\openk)\hskip 1truecm B : V\otimes V \rightarrow \openk\nn
B^\wedge(x\wedge y, z) &= \sum_{(z)}\,\pm\, B^\wedge(x,z_{(1)})B^\wedge(y,z_{(2)}) \nn
B^\wedge(x,y\wedge z) &= \sum_{(x)}\,\pm\, B^\wedge(x_{(1)},y)B^\wedge(x_{(2)},z) \, .
\end{align}
Since $B^\wedge$ is Laplace, it is a 2-cocycle and the circle product is associative. 
We know from \cite{fauser:2001b,brouder:fauser:frabetti:oeckl:2002a} that we can
distinguish two cases of such twists. If $B^\wedge$ is antisymmetric, then the twisted
algebra is isomorphic to the original algebra. $B^\wedge$ is a 2-coboundary in this case. 
However, the original grading remains only a filtration but can be newly established
with respect to the new product. That means we find in this case an isomorphism
\begin{align}
\Phi : V^{\wedge} &\rightarrow V^{\circ} \nn
\Phi(V^{\wedge^r}) &= V^{\circ^r}\oplus V^{\circ^{(r-2)}}\oplus \ldots \nn
\text{and}\hskip 0.5truecm V^{\circ} &= \sum_r V^{\circ^r} \,. 
\end{align}
This is the famous Wick expansion of quantum field theory \cite{fauser:2001b}.
If $B$ is symmetric, then the map is no longer an algebra isomorphism. The resulting
algebra is the Clifford algebra of the quadratic space $(V,Q)$, $Q(x)=B(x,x)$. Both
cases can be combined to come up with an arbitrary bilinear form. Our further objective
is to implement new tools to study these two cases of twist deformation.

\section{Singular value decomposition}

To be able to develop our new viewpoint, we need to address product and coproduct
maps as morphisms in $\textsf{mon}$. Hence we introduce a linearly ordered 
$\{e\}$-basis in $V^\wedge$ of dimension $2^{\textsf{dim} V}$. We consider form
now on the whole graded space $V^\wedge$ and this basis is linearly indexed. If
we focus on the generating space $V$, we will explicitely mention this. Using
this convention we obtain the maps
\begin{align}
m(e_i\otimes e_j) &= \sum_k m_{ij}^k e_k \,=\, \sum_k m_I^k e_k \nn
\Delta(e_i) &= \sum_{(e_i)} \,\pm\,\Delta_i^{kj} e_k\otimes e_j \,=\, 
\sum_{(e_K)} \,\pm\, \Delta_i^K e_K 
\end{align}
where $\{e_K\}$ is a linearly ordered basis of $V^\wedge\otimes V^\wedge$ and 
$\alpha$ the above defined (\ref{alpha}) encoding isomorphism $\alpha^{-1}(e_i\otimes e_j) 
= e_K$. It is obvious, that $m_I^k$ is a $4^n\times 2^n$-tensor, while $\Delta^I_k$ 
is an $2^n\times 4^n$-tensor.

To be able to derive invariant informations, like eigenvalues, we need to associate
quadratic matrices to $m$ and $\Delta$. We will concentrate on $m$, since $\Delta$ 
is treated analogously. Let $m^T$ denote the transposed matrix of $m$, \textit{ie} 
rows and columns interchanged. $m^T$ is a $2^n\times 4^n$-matrix . To be precise, 
in fact $m^T$ is the 
coproduct of the dual Hopf algebra $H^*$. To see this, let $\{f^i\}$ be the linearly 
ordered canonical dual basis of the $\{e\}$ basis. We have
\begin{align}
\textsf{eval}(f^i\otimes e_j) \,=\, f^i(e_j) &= \delta^i_j \nn
\textsf{eval}(f^i\wedge^\prime f^j \otimes e_k\wedge e_l) &=
\frac{1}{4}\textsf{eval}\big( (f^i\otimes f^j - f^j\otimes f^i) 
\otimes (e_k\otimes e_l - e_l\otimes e_k)\big) \nn
&= \frac{1}{2}(\delta^i_k\delta^j_l - \delta^i_l\delta^j_k)
\end{align}
etc. It is now possible to combine the morphisms $m$ and $m^T$ in two ways
\begin{align}
A = m\circ m^T &&& B = m^T\circ m \,, 
\end{align}
where $A$ is a $2^n\times 2^n$-matrix and $B$ is a $4^n\times 4^n$-matrix. Both 
matrices are symmetric by construction and can be diagonalized by an unitary 
(orthogonal) matrix ($U : V^\wedge \rightarrow V^\wedge$, and $
V : (V^\wedge\otimes V^\wedge)\rightarrow (V^\wedge\otimes V^\wedge)$)
\begin{align}
D_A = UAU^T &&& D_B = VAV^T \nn
U\,U^T = \Id_V &&& V\, V^T = \Id_{V\otimes V}\,.
\end{align}
Since $A^T=A$ and $B^T=B$ are non-negative matrices, we can compute the square 
root of $D_A$ and $D_B$ using functional calculus. We denote by $D_A^{\frac{1}{2}}$ 
the $n\times m$ respectively $(D_A^{\frac{1}{2}})^T$ the $m\times n$ matrices which 
square to $D_A$ and $D_B$. For ease of notation we drop the transposition ${}^T$ since
the shape of $D_A^{\frac{1}{2}}$ is obviouis from the context. This allows us to write
\begin{align}
A &= U^T D_A^{\frac{1}{2}} D_A^{\frac{1}{2}} U \nn
  &= U^T D_A^{\frac{1}{2}} V\, V^T D_A^{\frac{1}{2}} U \,=\, m \circ m^T \nn
B &= V^T D_B^{\frac{1}{2}} U\, U^T D_B^{\frac{1}{2}} V \,=\, m^T \circ m \, .
\end{align}
Therefore one concludes that $D_A\oplus \textsf{ker}(m) = D_B$ and especially that
the \textit{set} of positive eigenvalues of $A$ and $B$ are identical. The
eigenvalues of $D_A$ or $D_B$ are called singular values, they are positive by 
construction. One has to be careful during the identification of the two maps, 
since they agree only up to isomorphism (a permutation of the singular values). 
However we can state the following, now obvious, theorem, which was stated originally 
by Oziewicz using laborious computations:

{\bf Theorem [Oziewicz \cite{oziewicz:2001a}]} The operators $m\circ \Delta$ and 
$\Delta\circ m$ fulfill the same minimal polynomial and differ only in the dimension
of their kernels. 

This is quite important, since it is also a statement about the right-hand-side
of eqn. (\ref{AXIOM1}), a fundamental axiom of bi- and Hopf algebras.

Let now $\{u_i\}$ be the set of column vectors of $U$ and $\{v_I\}$ be the set
of column vectors of$V$ and let $\{d_i\}$ be the set of positive singular values
of $D_A$ or $D_B$. It is now possible to relate the two sets of vectors via
\begin{align}
m\, v_I 
&= \pm (d_A^{\frac{1}{2}} )_{ \alpha^{-1}([i,1]) } \, u_{ \alpha^{-1}([i,1]) } 
\,\cong\, \pm (d_A^{\frac{1}{2}} )_i \, u_i 
\nn
m^T\, u_i \,\cong\, m^T\, u_{\alpha(I)}
&= \pm (d_B^{\frac{1}{2}})_I\, v_I \,.
\end{align}
Using that particular isomorphism $\alpha$ which relates the index sets $\{I\}$ 
and $\{i\}$ in such a way that $\alpha^{-1}(I)=[i,1]$ picks eigenvectors to the same 
singular value $d^{\frac{1}{2}}_i$ One can then come up with a spectral 
decomposition of the product and coproduct maps. Our choice of the signs in the 
square roots fixes the maps completely. Note that eigenvectors are assumed to 
be nonzero, orthogonal and normalized. However, from $v_i\cdot v_i =1$ we can 
fix $v_i$ only up to sign. We may hence choose positive signs, finding
\begin{align}
m   &= \sum_i u_i (d_A^{\frac{1}{2}})_i v^T_{\alpha^{-1}([i,1])} \nn
m^T &= \sum_i v_{\alpha^{-1}([i,1])} (d_A^{\frac{1}{2}})_i u^T_i \, .
\end{align}
where the sum is over all positive singular values. 

\section{Singular value decomposition for Grassmann and Clifford algebras}

\subsection{Grassmann case:}

We proceed to calculate explicitly the singular values for Grassmann  and Clifford
algebra products and coproducts of course. We start with the Grassmann case and
compute the $d_i$ for the composition $A=m\circ \Delta$. Therefore we note, that the 
coproduct of a basis element $e_{i_1}\wedge \ldots\wedge e_{i_r}$ is given by
all $(p,q)$-shuffles of the indices $(i_1,\ldots,i_r)$, where $p+q=r$. Wedging
each of these terms back, one obtains the original basis element. Hence we find
\begin{align}
m \circ \Delta(e_{i_1}\wedge \ldots\wedge e_{i_r}) &= \#\text{of $(p,q)$-shuffles~}
\,\cdot\, (e_{i_1}\wedge \ldots\wedge e_{i_r}). 
\end{align} 
To compute the number of $(p,q)$-shuffles with $p+q=r$, we need to select 
zero, one, two, etc elements out or $r$ elements, getting $r$ choose $p$ such 
sequences. Summing up, we get $2^r$ terms. If we introduce the grade operator
$\partial$ as
\begin{align}
\partial : \sum V^\wedge \rightarrow \openN &&& \partial V^{\wedge^r} = r 
\end{align}
we can write our result as

{\bf Theorem [Oziewicz \cite{oziewicz:1997a}]} The operator $A=m^\wedge \circ 
\Delta^\wedge$ acts as the linear operator $2^\partial$ on $V^\wedge$.

This is a well known result, but we can generalize this in the following way

{\bf Theorem} The operator $A^{(r)} = (m^\wedge)^{r-1} \circ (\Delta^\wedge)^{r-1}$
acts as linear operator $r^\partial$ on $V^\wedge$ ($A=A^{(2)}$, $A^{(1)}=\Id$). 

{\sc Proof:} We need to count the number of $(p_1,\ldots,p_r)$-shuffles with 
$\sum p_i=\textsf{dim}V$, related to multinomial coefficients, while we had to 
count binomial coefficients in the preceding theorem.

Note that these operators are homogeneous with respect to the grade and commute
with the grade operator $\partial$. Hence they are constant on each space 
$V^{\wedge^r}$. Hence we cannot drop smaller eigenvalues since all of them are
equal on all homogeneous elements. However, the higher grade elements have larger
singular values. A map $F : V^{\wedge} \rightarrow V^{\wedge}$ can however be
considered to have more weight on the higher grade subspaces.

Knowing the singular values, we can easily write down the minimal polynomial of the 
operators $A$ and $A^{(r)}$
\begin{align}
\prod_{i=0}^{\textsf{dim}V} (A - 2^i)=0 &&&
\prod_{i=0}^{\textsf{dim}V} (A^{(r)} - r^i)=0 
\end{align}
The geometric degeneracies of the eigenspaces are given by binomial and multinomial 
coefficients and we can infer the characteristic polynomials too, \textit{eg}
\begin{align}
\prod_{i=0}^{\textsf{dim}V} (A - 2^i)^{\textsf{dim}V \choose i}&=0 \nn
B^{(4^{\textsf{dim}V}-2^{\textsf{dim}V})}\,
\prod_{i=0}^{\textsf{dim}V} (B - 2^i)^{\textsf{dim}V \choose i}&=0 \, .
\end{align}

The grade operator applied directly to the index set returns simply the 
cardinality of the index set $\vert\{i_1,\ldots,i_r\}\vert=r$. The Grassmann product 
and coproduct maps have therefore the spectral decomposition
\begin{align}
m &= \sum_{i=1}^{{\textsf{dim}V}}
u_i\,\, 2^{\frac{1}{2}\,\vert \alpha^{-1}([i,1])\vert}\,\, v^T_{\alpha^{-1}([i,1])} \nn
m^T =\Delta &= \sum_{i=1}^{{\textsf{dim}V}}
 v_{\alpha^{-1}([i,1])}\,\, 2^{\frac{1}{2}\,\vert i\vert}\,\, u^T_i \, ,
\end{align} 
where the sum is over all nonzero singular values and $\alpha$ is the particular 
index isomorphism which guarantees that $u_i$ and $v_{\alpha^{-1}([i,1])}$ belong
to the same singular value.

\subsection{Clifford case:}

The Clifford case is much more involved. We can distinguish three cases. Either
deform the product, the coproduct or both. Since we use ordinary transposition 
to obtain $m^T=\Delta$, hence identifying the Hopf algebras $H$ and $H^*$, we 
cannot do this independently, unless we allow nontrivial dual isomorphisms. In
this case, the dual basis is given by $f^i(e_j) = h^i_j$ where $h^i_j$ is a $GL(n)$
element. While this may be of importance in geometry and physics, see 
\cite{fauser:stoss:2004a}, we will not include here this complication. We will use
product deformations, and the coproduct is deformed by demanding an Euclidean 
duality isomorphism, \textit{ie} $\delta^i_j$.

However, we will allow deformations by symmetric or nonsymmetric bilinear forms. 
We will postpone the general case to the computer algebra experiment, and concentrate
here on the following. Let $g: V\otimes V \rightarrow \openk$ be a symmetric non 
degenerate bilinear form. Let $\Delta(g)=g_{(1)} \otimes g_{(2)}$. We 
define the twisted (Clifford or circle) product and coproduct maps as
\begin{align}
m_g(x\otimes y) \,=\, x\circ_{g} y &= \sum_{(x),(y)} 
(-1)^{\partial x_{(2)}\partial y_{(1)}}\,g^\wedge
(x_{(1)},y_{(1)})   x_{(2)}\wedge y_{(2)} \nn
\Delta_{g^{\prime}}(x) &= \sum_{(x)}
(-1)^{\partial g^{\prime}_{(2)} \partial x_{(1)}}
 g^{\prime}_{(1)}\wedge x_{(1)} \otimes   g^{\prime}_{(2)}\wedge x_{(2)}
\end{align}
The coproduct with respect to the metric $g$ can be written as
\begin{align}
\Delta_g(x) &= \Id\otimes \Id + \sum_{i,j} g_{ij}x_i\otimes x_j
+ \sum_{i<j, k<l}\frac{1}{2!}(g_{ik}g_{jl}-g_{il}g_{jk}) 
x_i\wedge x_j \otimes x_k\wedge x_l +\ldots
\end{align}
where the decomposable element $x$ is given as a monomial in the $x_i$ and the 
expression is extended by linearity to $V^\wedge$, see \cite{fauser:2002c}. 
From the preceding two expressions we deduce, that the coproduct $\Delta$ 
obtained by transposition of the multiplication table $m_{ij}^k$ is given by
the deformation w.r.t. the numerically identical cometric $g^{\prime}$, 
\textit{ie} we have $g\equiv g^{\prime}$. 

{\bf Example} Let $\textsf{dim}V =1$ and introduce the metric $g(e_1,e_1)=a$. We use
the basis $\{\Id=e_0,e_1\}$ for $V^\wedge$ and $\{\Id\otimes\Id,e_1\otimes\Id,
\Id\otimes e_1,e_1,\otimes e_1\}$ for $V^\wedge\otimes V^\wedge$, with shorthand
$\{e_{0,0},e_{1,0},e_{0,1},e_{1,1}\}$. We find the multiplication table and the 
section coefficients (comultiplication table)
\begin{align}
m_g\cong \begin{array}{c|cccc}
m_g   & e_{0,0} & e_{1,0} & e_{0,1} & e_{1,1} \\\hline
e_0 &    1    &     0   &    0    &    a    \\
e_1 &    0    &     1   &    1    &    0
\end{array}
&&&
m_g^T\cong \begin{array}{c|cc}
m_g^T   & e_0 & e_1 \\\hline
e_{0,0} &    1    &     0 \\
e_{1,0} &    0    &     1 \\
e_{0,1} &    0    &     1 \\
e_{1,1} &    a    &     0
\end{array}
\end{align}
The matrices $A=m_g\circ m_g^T$ and $B=m_g^T\circ m_g$ read then
\begin{align}
A \cong \left[\begin{array}{cc} 1+a^2 & 0 \\ 0 & 2 \end{array}\right] \nn
B \cong \left[\begin{array}{cccc}
1 & 0 & 0 & a \\
0 & 1 & 1 & 0 \\
0 & 1 & 1 & 0 \\
a & 0 & 0 & a^2
\end{array}\right]
\end{align}
The eigenvalues are hence $1+a^2$, $2$ leading to the singular values $\sqrt{1+a^2}$,
$\sqrt{2}$. The matrix $A$ is already diagonal, showing that $e_0,e_1$ are 
orthonormalized eigenvectors $\{u_i\}$. However, we need to orthogonalize $B$. We
can arrange the new basis $\{v_i\}$ as
\begin{align}
\lambda&=(1+a^2) &&:& v_1&=\frac{1}{\sqrt{1+a^2}}(\Id\otimes \Id + a\,e_1\otimes e_1) \nn
\lambda&=2       &&:& v_2&=\frac{1}{\sqrt{2}} (e_1\otimes \Id+ \Id\otimes e_1) \nn
\lambda&=0       &&:& v_3&=\frac{1}{\sqrt{2}} (e_1\otimes \Id- \Id\otimes e_1) \nn
\lambda&=0       &&:& v_4&=\frac{1}{\sqrt{1+a^2}}(a\,\Id\otimes \Id - e_1\otimes e_1)
\end{align}
Note, that the product map acting on the $v_i$ yields the square root of the singular 
values times the column (eigen)vectors $u_i$. Especially $m(v_3)=0$ and $m(v_4)=0$, showing
that $\textsf{ker}(\, m\,) \cong \textsf{lin-hull}(v_3,v_4)$. The product and coproduct spectral
decompositions are given as
\begin{align}
m(x\otimes y) &= \sum_{i=1}^2 u_i d^{\frac{1}{2}}_i v^T_i(x\otimes y) \nn
&= \Id \sqrt{1+a^2} \frac{1}{\sqrt{1+a^2}}
\Big(\Id(x)\otimes \Id(y) + a\,e_1(x)\otimes e_1(y)\Big)\nn
&~~~~
+ e_1 \sqrt{2} \frac{1}{\sqrt{2}}  
\Big(e_1(x)\otimes \Id(y)+ \Id(x)\otimes e_1(y)\Big)   \nn
&= g(\Id,x)g(\Id,y) + a\,g(e_1,x)g(e_1,y) \nn
&~~~~
+ e_1\big(g(e_1,x)g(\Id,y)+g(\Id,x)g(e_1,y)\big) \\
m^T(x) &=  \sum_{i=1}^2 v_i d^{\frac{1}{2}}_i u^T_i(x) \nn
       &= (\Id\otimes \Id + a\,e_1\otimes e_1)g(\Id,x) 
         +(e_1\otimes \Id- \Id\otimes e_1)g(e_1,x)
\end{align}
Setting $a=0$ returns the Grassmann case. The particular choice $a=\pm i$, the 
complex number unit, increases the degeneracy and one has a three dimensional null 
space. Note that one eigenvalue is equal to $\textsf{dim}V^\wedge=2$, but that 
the other one depends in general on $a$. This illuminates the following 

{\bf Theorem [Oziewicz]} If $m_g$ is twisted by a symmetric
nondegenerate bilinear form $g$ and $\Delta_{g^{-1}}$ is deformed by $g^{-1}$
then the operator $A=m_g\circ \Delta_{g^{-1}}$ acts as the multiplication by 
$\textsf{dim}V^\wedge$.

This theorem can readily be generalized.

{\bf Theorem} If $m_g$ is twisted by a symmetric nondegenerate bilinear form
$g$ and $\Delta_{g^{-1}}$ is twisted by $g^{-1}$, then the operators
$A^{(r)}= m_g^{r-1}\circ \Delta_{g^{-1}}^{r-1}$ acts as multiplication operators
$(\textsf{dim}V^\wedge)^{r-1}$, in particular $A^{(2)}\cong\textsf{dim}V^\wedge$.

{\sc Proof:} A trivial iteration of the preceding theorem.

In particular our outcome shows directly that the condition that the deformations 
are mutually related via the inverse metrics is necessary.

{\bf Theorem [Oziewicz]} \label{OziThrm} If the cliffordization is performed 
w.r.t. a (symmetric) metric $g$ and the coproduct is deformed w.r.t. the 
cometric $g^{-1}$, such that $g g^{-1}=\Id$, then the convolution has no antipode.

This result renders the codeformation w.r.t. the inverse to be a particular 
singular and unuseful situation if a pseudoinverse (antipode) is needed. 
Especially in physics a pseudoinverse is however desirable in most cases.
A way out of this problem was investigated in \cite{fauser:oziewicz:2001a}.

In our case, since we had demanded that $\Delta=m^T$, we obtain the singular case
for symmetric metrics fulfilling $g^2=\Id$. All these matrices are in the orbit
of diagonal matrices with diagonal entries $\pm 1$. Due to Sylvester's theorem, all 
$GL(n,\openk)$ matrices fall into an orbit of such an element, as long as the
ground field $\openk$ is of characteristic $0$. In particular, $GL(n,\openC)$
has one such orbit while $GL(n,\openR)$ hast $n+1$ such orbits, characterized
using the signature. This is related to the Brauer-Wall group of quadratic forms, 
see \cite{hahn:1994a}. Hence after normalizing $g(e_1,e_1)=a=1$ we find in our 
above example Oziewicz's result. However, spin groups or special orthogonal
groups, as symplectic groups do not allow such a rescaling. For a brief relation
of this outcome to group branching laws see section \ref{applications}.

Let us deviate a little bit from Clifford topics and consider the group like 
coproduct $\delta(x)=x\otimes x$ for all $x$. One can show, that the pair
of morphisms $m^\wedge,\delta$ still fulfills the axiom (\ref{AXIOM1}), but
that in general for this and twisted such products no antipode exists. Dualizing 
this time the comultiplication, results in a product map $\delta^T=m^B$. This
product turns $V$ into a Boolean algebra (all elements are idempotent) 
\begin{align}
m^B(x\otimes y) &= \left\{\begin{array}{cl}
x & \text{if $x=y$} \\
0 & \text{otherwise}
\end{array}\right.
\end{align} 
The matrix $A=m^B\circ \delta$ is the unit matrix in $\textsf{dim}V$ dimensions
and $B=\delta\circ m^B$ is a diagonal matrix with $\textsf{dim}V$ ones and
zeros otherwise. A twist deformation in this case transforms the elements from
being idempotent to being almost idempotent, hence an uninteresting map.
However, disregarding the transposition as being Euclidean, we can combine
$m^\wedge\circ \delta$, which is related to inner products of group 
representations and nontrivial. Note that for group like situations we obtain
full degeneracy. Hence the classical geometric case is characterized by
total degeneracy of the product and coproduct maps.

Now, let us assume that we have a symmetric bilinear form $g$. It is possible
to diagonalize this form in the space $V$\!, we can deduce then $g^\wedge$ 
and obtain for the matrix $A=m_g^T\circ m_g$ the diagonal representation
\begin{align}
g &: V\otimes V \rightarrow \openk 
\hskip 1truecm g\,=\,\textsf{diag}(l_1,\ldots,l_n) \nn
g^\wedge &: V^\wedge\otimes V^\wedge \rightarrow \openk \nn
g^\wedge &=
\textsf{diag}(L_0,2^{n \choose 1} L_{i}^{n \choose 1}, 
                  2^{n \choose 2} L_{ij}^{n \choose 2},
          \ldots, 2^{n \choose {n-1}} L_{i_1,\ldots,i_{n-1}}^{n \choose {n-1}}, 
                  2^{n \choose n} )
\label{evals}
\end{align}
where we have split off the Grassmann eigenvalues $2^{n \choose m}$, and the 
metric dependent parts $L_{i_1\ldots}$. Superscripts of the $L_{i\ldots}$ denote 
the 'degeneracy' of eigenvalues of the same type but different index structure. 
The subscripts denote which indices are missing in the total index set 
$\{1,\ldots,\textsf{dim}V\}$. The $L_{i_1,\ldots,i_r}$ read as follows, where
the sums range over in $\{1,\ldots,n\}$ omitting $\{i_1,\ldots,i_r\}$ which index
the basis of $V^\wedge$. It is clear that there are $n \choose r$ such sets which
explains the `degeneracy`. Hence we find 
\begin{align}
L_{0}&=
1+\sum_{i_1} l^2_{i_1} +\sum_{i_1<i_2}l^2_{i_1}l^2_{i_2} 
+ \sum_{i_1<i_2<i_3}l^2_{i_1}l^2_{i_2}l^2_{i_3}+ \ldots 
+ \sum_{i_1<\ldots<i_{n-1}}l^2_{i_1}\ldots l^2_{i_{n-1}}\nn
L_{i}&=
1+\sum_{i_1} l^2_{i_1} +\sum_{i_1<i_2}l^2_{i_1}l^2_{i_2} 
+ \sum_{i_1<i_2<i_3}l^2_{i_1}l^2_{i_2}l^2_{i_3}+ \ldots 
+ \sum_{i_1<\ldots<i_{n-2}}l^2_{i_1}\ldots l^2_{i_{n-2}}\nn
L_{ij}&=
1+\sum_{i_1} l^2_{i_1} +\sum_{i_1<i_2}l^2_{i_1}l^2_{i_2} 
+ \sum_{i_1<i_2<i_3}l^2_{i_1}l^2_{i_2}l^2_{i_3}+ \ldots 
+ \sum_{i_1<\ldots<i_{n-3}}l^2_{i_1}\ldots l^2_{i_{n-3}}\nn
L_{ijk}&=
1+\sum_{i_1} l^2_{i_1} +\sum_{i_1<i_2}l^2_{i_1}l^2_{i_2} 
+ \sum_{i_1<i_2<i_3}l^2_{i_1}l^2_{i_2}l^2_{i_3}+ \ldots 
+ \sum_{i_1<\ldots<i_{n-4}}l^2_{i_1}\ldots l^2_{i_{n-4}} \nn
&~
\ldots\nn
L_{i_1,\ldots,i_r}&=
\sum_{s=0}^r \sum_{i_1<\ldots<i_s} l^2_{i_1}\ldots l^2_{i_{n-s-1}} \nn
&~
\ldots\nn
L_{i_1,\ldots,i_n}&=1
\end{align}
The sum over an empty index set is defined to be $1$. It is obvious from the 
form of the $L_{i,\ldots}$, that the degernaracy of the eigenvalues is in general
removed if the $l_i$ are mutually different. These functions are related to 
elementary symmetric functions in the variables $l_i^2$ where $i$ runs in $\{1,
\ldots,\textsf{dim}V\}$ with $\{i,\ldots\}$ omitted in $L_{i,\ldots}$. 

As special cases we notice, that for selfinverse metrics $g\circ g =\Id$
we need to have $l_i=\pm1$, and hence $l^2_i=1$. The eigenvalues are then 
given by the number of the terms in $L_i$ times the Grassmann eigenvalues. This
recovers Oziewicz's theorem that $A$ is fully degenerate with eigenvalues
$\textsf{dim}V^\wedge = 2^{\textsf{dim}V}$. A second special case is $l_i=0$ 
for all $i$ which reduces to the Grassmann case. 

Let now $f : V\otimes V \rightarrow \openk$ be a totally antisymmetric bilinear 
form and extend it as above via Laplace expansion to $f^\wedge$. As a consequence,
we see that the $\{u_i\}$ basis is no longer an eigenbasis to $A= m_f^T\circ m_f$.
The new eigenbasis introduces a $f$-dependent filtration of the algebra. This 
filtration can be turned into a gradation which was described by dotted wedge 
products in previous works
\cite{fauser:1996c,fauser:2001e,fauser:ablamowicz:2000c,fauser:2002c}. Exactly
this new filtration establishes the Wick reordering of quantum field theory
\cite{fauser:2001b}. Hence a basis transformation in $V^\wedge$, acting as 
identity of $V$ however, establishes the new gradation. A spectral decomposition 
of the product map has to use this new basis w.r.t. the newly established
$f$-grading.

We know from cohomological considerations \cite{brouder:fauser:frabetti:oeckl:2002a}, 
that antisymmetric and symmetric twists fall into two classes, namely proper
2-cocycles and 2-coboundaries. This explains their different algebraic behaviors
and allows to study the two cases independently. The general case is a convolutional
mixture of these two possibilities. From group theory we know, that introducing a
2-cocycle might make it necessary to come up with the need of a change in the
filtration of the algebra \cite{fauser:jarvis:2003a}.

Finally, we might remark, that the singular value decomposition allows to
provide estimates on certain norms of the operators under consideration. 
The Frobenius norm of a $n\times m$-map $m$ is defined as
\begin{align}
\sum_{i,j} m_{ij}^2 &= \sum_k (d^{\frac{1}{2}}_k)^2
\end{align}
while the operator 2-norm is given as
\begin{align}
\vert\vert m\vert\vert_2 &= \sup_{\vert v\vert =1} \vert m v\vert 
    \,=\, d^{\frac{1}{2}}_1
\end{align}
where $d^{\frac{1}{2}}_1$ is the greatest singular value. In particular, we note 
that the Clifford and Grassmann multiplication maps are unbounded operators
if $\textsf{dim}V$ goes to infinity, \textit{eg} is an $L^2$ space. The grow is
exponential and the divergence hence serious.

\subsection{Spectral form of product coproduct pairs}

Now, let $m$, $\Delta=m^T$ be a product coproduct pair related by the Euclidean
dual isomorphism, \textit{ie} via transposition. Let $A=m\circ\Delta$ be the
associated symmetric operator $A : V^\wedge \rightarrow V^\wedge$ with canonically 
normalized eigenvector basis $\{u_i\}$, $Au_i=\lambda_iu_i$. The $\{u_i\}$ form the 
column vectors of the matrix $U$ which diagonalized $A$. Let $B=\Delta\circ m$, a 
symmetric operator, $B : V^\wedge\otimes V^\wedge \rightarrow V^\wedge\otimes V^\wedge$,
having canonically normalized column eigenvectors $\{v_I\}$, which form the column
vectors of the matrix $V$ diagonalizing $B$. We can summarize our findings in 
the following

{\bf main theorem:} The coproduct $m^T=\Delta$ maps the column eigenvectors $u_i$ of 
$A$ onto the column eigenvectors $v_i$ of $B$ w.r.t. the same singular value and 
vice versa maps the product the $v_i$ onto the $u_i$. Let the canonical normalization
be $UU^T=D_A$ and $VV^T=D_B$. Then the product has the spectral
form
\begin{align}
m &= \sum_{i} u_i \,\,\Delta(u_i)^T 
\end{align}
and the coproduct has the spectral form
\begin{align}
\Delta\,=\, m^T &= \sum_{\{I \vert m(v_I)\not=0\}} v_I\,\, m(v_I)^T \, .
\end{align}

This amazing result technically allows to math the eigenvectors $\{u_i\}$
and $\{v_I\}$ via the Hopf algebra structure, since the coproduct exactly 
matches pairs of eigenvectors for a particular singular value. The 
computational technicallity of matching eigenvectores is hence resolved.
Furthermore, the computation of the eigenvectors $\{u_i\}$ of $A$ is 
considerably simpler than that computation of the eigenvectors $\{v_I\}$ 
for $B$, which can now be obtained from the application of the coproduct
directly. The operators $A$ and $B$ are easily derived in spectral form 
as
\begin{align}
A &= m_g\circ m^T_g \,=\, 
  \sum_{i,I} u_i \otimes (\Delta_g(u_i)^T\mid v_I) \otimes m(v_I)^T \nn
B &= m_g^T\circ m^T \,=\, 
  \sum_{i,I} v_I \otimes (m(v_I)^T \mid u_i) \otimes \Delta(u_i)^T
\end{align}
which holds true for \textit{any} basis of $A$ and $B=A\otimes A$ of course.

We mention here explicitely the technical importance of this result. As discussed
in the introduction, SVD is a powerful and widely used tool for data compression,
analysis of data, searching, image processing etc. A Hopf algebraic point of view,
employing the computational accessible coproduct, saves lots of computation time
and even bandwidth in transmitting data, since only the $\{u_i\}$ eigenvectores, and 
the singular values have to be sent, since the much more involved $\{v_I\}$
follow \textit{uniquely} from the coproduct structure. Technical applications
are based on the case studied in this paper, where product and coproduct are
related by Euclidean duality, \textit{ie} via transposition. In fortunate situations
the coproduct may be known, and no information about it has to be transmitted
at all. If the space $A$ is graded, the information of the coproduct is reduced
to the action on the grade 1 space and expanded using the homomorphism property
eqn. (\ref{AXIOM1}). Of course, images may not have a product coproduct structures
in general, so care is needed. However, see \cite{ablamowicz:2002a} for an 
embedding of matrix SVD into a Clifford algebraic setting.

\section{CAS experiment in dimension 2}

Since its a difficult task to compute the singular values, vector space
decompositions etc in the general twisted case, we consider here 
$\textsf{dim}V=2$ and use a computer algebra system (CAS) to solve the 
general setting for an arbitrary suitably chosen bilinear form $B$.
We use CLIFFORD and BIGEBRA \cite{clifford,BIGEBRA} packages for
Maple \cite{MAPLE} \footnote{%
A Maple worksheet containing the computations of this section is available 
from the author or from the url:
{\tt\small http://clifford.physik.uni-konstanz.de/\~{}fauser/}.}.

Since we are mainly interested in a model which allows a physical interpretation,
we choose the metric
\begin{align}
B &= \left[\begin{array}{cc}
0 & \rho+\nu \\
\rho-\nu & 0
\end{array}\right]\,.
\end{align}
The commutation and anticommutation relation for the ${e_i}$ follow as
\begin{align}
\begin{array}{c|cccc}
\{.\mid.\}_+ & e_0 & e_1 & e_2 & e_{12} \\ \hline
e_0    & 2\Id   & 2e_1            & 2e_2            & 
    2e_{12} \\
e_1    & 2e_1   &  0              & 2\rho\Id         & 
    -2\nu e_1 \\
e_2    & 2e_2   & 2\rho\Id         & 0               & 
    -2\nu e_2 \\
e_{12} & 2e_{12}& -2\nu e_1 & -2\nu e_2 & 
     2(\rho^2-\nu^2)\Id-4\nu e_{12}
\end{array}
\end{align}

\begin{align}
\begin{array}{c|cccc}
{}[.\mid.]_- & e_0 & e_1 & e_2 & e_{12} \\ \hline
e_0 & 0 & 0 & 0 & 0 \\
e_1 & 0 & 0 & 2\nu\Id+2e_{12}   & -2\rho e_1 \\
e_2 & 0 & -2\nu\Id-2e_{12} & 0  &  2\rho e_2 \\
e_{12} & 0 & 2\rho e_1             & -2\rho e_2 & 0 
\end{array}
\end{align}

\begin{figure}
\hfil
\includegraphics[width=6.0cm]{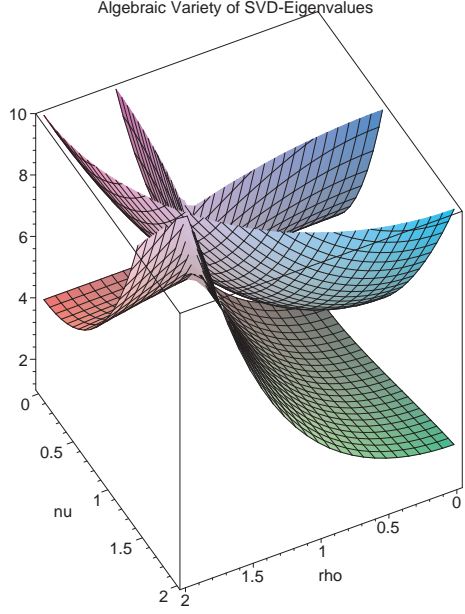}
\hfil
\caption{Eigenvalue surfaces over the $\rho$-$\nu$-plane. Shown is
one quadrant, the other three are mirror symmetric w.r.t. the 
$xz$- and $yz$-planes. Remarkable is that all three planes meet 
in a one-dim. curve.}
\label{fig1}
\end{figure}

It is obvious that with the identification $a=e_1$, $a^\dagger=e_2$ we find that
the CAR relations
\begin{align}
\{a,a^\dagger\}_+ &= 2\,\rho \Id
\end{align}
hold. For a detailed discussion of this and a 4-dimensional model see
\cite{fauser:2001e}. The multiplication table is given as
\begin{align}
m_B\,=\,\hskip 0.7\textwidth\nn
 \left[\begin{array}{cccccccccccccccc}
1&0&0&0& 0 &0&\rho-\nu&0           &0&\rho+\nu&0&0       &0
    & 0 & 0 & \rho^2-\nu^2 \\
0&1&0&0& 1&0       &0     &\rho-\nu&0&0      &0&0       &0
    &-\rho-\nu & 0& 0 \\
0&0&1&0& 0&0           &0           &0&1      &0&0&-\rho-\nu
    &0 & 0 &\rho-\nu & 0 \\
0&0&0&1& 0&0          &-1           &0&0      &1&0       &0
    &1 & 0 & 0 & -2\nu
\end{array}\right]
\end{align}
\begin{figure}
\hfil
\includegraphics[width=6.0cm]{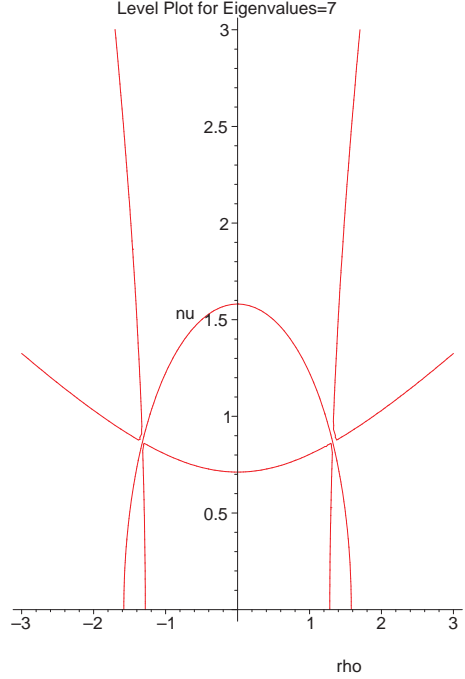}
\hfil
\caption{Cross section for $z=7$ of figure \ref{fig1}. The plot
shows clearly the confocal level crossing of the eigenvalue surfaces.}
\label{fig2}
\end{figure}
The matrix $A = m\circ m^T$ is hence given as
\begin{align}
A&=m_B\circ m_B^T\,=\,\left[\begin{array}{cccc}
a & 0 & 0 & b \\
0 & c & 0 & 0 \\
0 & 0 & c & 0 \\
b & 0 & 0 & d
\end{array}\right] \\
a&= (\nu^2+1+2\rho\nu+\rho^2)(\nu^2+1-2\rho\nu+\rho^2) &&&
b&= 2\nu(1-\rho^2+\nu^2) \nn
c&= 2+2\rho^2+2\nu^2 &&& d&= 4+4\nu^2
\end{align}
We can identify the following special cases: 
\begin{itemize}
\item $\rho=0$ is the $\nu$ dependent Grassmann case. However, even 
in this case the deformed algebra obeys a new filtration, which is 
imposed by $\nu$. 
\item $A$ is diagonal, if $b=0$, from which follows: 
$\rho=\pm\sqrt{1+\nu^2}$ or $\nu=0$. The eigenvalues are in this case of Clifford
type and the fourfold degenerated eigenvalues are $4+4\nu^2$.
\end{itemize}

A remarkable fact is displayed in Figure \ref{fig1}. All three\footnote{%
Actually four surfaces, but two surfaces are degenerate., see eqn (\ref{evals}).} 
eigenvalue surfaces, emerging from the three types of rank, 0,1, and 2, meet 
in a single curve. This curve will be called singular locus, since it establishes 
a relation between $\rho$ and $\nu$ in such a way that all eigenvalues are 
degenerated. In fact, this situation is singular in a much more peculiar way. 
The relation imposed, $\rho=\pm\sqrt{1+\nu^2}$, also implies that the metric 
$B$ on $V$ squares (as a matrix) to one. Therefore, the coproduct is based on
$B^{-1}$ and the theorem of Oziewicz (see page \pageref{OziThrm}) stating that
no  antipode can exist applies in this case. In Figure \ref{fig1}, we display the 
positive $\rho$-$\nu$-quadrant of the algebraic varieties defined by the
eigenvalues. The other quadrants are obtained by mirroring through the $xz$- 
and $yz$-planes. Two surfaces are saddle shaped, one has a (higher order) parabolic
form. The incidence of all three surfaces is obvious from this plot.

In Figure \ref{fig2}, we plot a section for constant $z$-value (\textit{ie} $z=7$).
It is clearly seen how the surfaces intersect in a single curve (point in this
section). Seen as eigenvalues, a \textit{level-crossing} takes place, which
is not correctly displayed in the plot, due to the contour plot option of 
Maple. One surface is doubly degenerated, since the matrix $A$ has 4 
eigenvalues, but due to the grading in our setting only three of them are 
different. 

\begin{figure}
\hfil
\includegraphics[width=6.0cm]{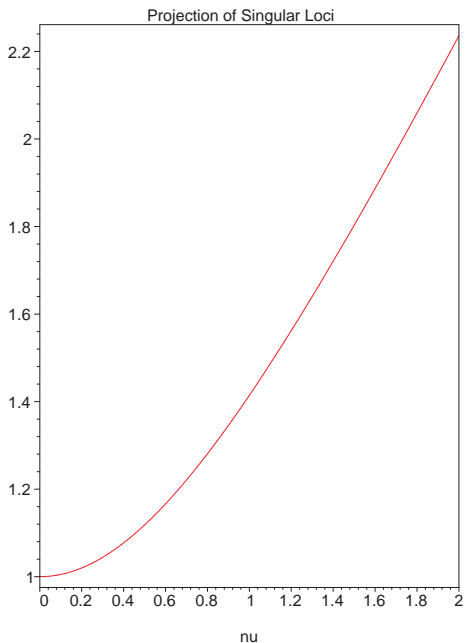}
\hfil
\caption{Plot of the singular curve $\prod_{i\not= j} (L_{i}-L_{j})=0$, 
\textit{ie} $\rho=\sqrt{1+\nu^2}$ of maximal degeneracy.}
\label{fig3}
\end{figure}

The commutation relations used in physics, having $\rho=\hbar/2$, does, in 
units of $\hbar$, \textit{not} reach the degenerate case. This makes a 
difference only, if one assumes that a rescaling is not possible. Hence, if
we agree that we have (half) integral values for $\hbar$, measured in units of
$\hbar$, we need to assume \textit{higher spin values} to be realized to reach
degeneracy. Since $\nu$ is not quantized, it can be arranged to hit the degeneracy,
but only for sufficient large $\rho$. This correlation is displayed in Figure 
\ref{fig3}. We plot there the projection of the singular curve into the 
$\rho$-$\nu$-plane. Its easily seen that singularities need $\rho>1$ to occur,
that the asymptotics is $\rho(\nu)=1$ for $\nu\rightarrow 0$ and 
$\rho(\nu)\simeq\nu+\text{const}$ for $\nu \rightarrow \infty$. 

\section{\label{applications}Connection to other applications}

\subsection{Symmetric functions, Schur functors}

During the work on symmetric functions \cite{fauser:jarvis:2003a} it became 
clear, that the homomorphism axiom, see eqn. (\ref{AXIOM1}), is equivalent to 
group branching  rules. Our results on singular values suggest, that the split 
into degeneracy subspaces can be described by methods from invariant theory. In 
this sense, one can assume that the spaces are direct sums and carry a (quantum) 
group action. More over, the eigenvalues should then have a combinatorial 
interpretation and it should become possible to compute them in a more effective 
way. Hence looking in two different ways at the decomposition 
$U(n) \downarrow U(n)\otimes U(n)$
via product-coproduct or product-coproduct maps allows to connect the representations
of the two sides also. Hence SVD is a Glebsch-Gordan problem in disguise.
Knowing the branching rules is hence connected with knowing the spectral decomposition
of the product and coproduct maps.

Classical invariant theory uses Schur functions to describe invariant subspaces.
This method can be generalized to the functorial level where Schur functors
characterize invariant subspaces as such, not supporting a basis. The main
point is, that Schur functions allow, via the Littlewood-Richardson rule,
the evaluation of the product $V_\lambda\otimes V_\mu = 
\sum_{\nu} c^\nu_{\lambda\mu}V_\nu $. In \cite{fauser:jarvis:2003a} it was
schown, how the cohomological Hopf algebra approch helpes to understand 
group branching laws. The SVD is hence connected to a direct computation
of the invariant subspaces. This can be achieved by introducing new types
of coproducts. \textit{Eg} we can pick involutions $\sigma$ in $V$ and define
a new coproduct $\Delta_\sigma = (\sigma\otimes \sigma)\circ\Delta\circ \sigma$,
which \textit{cannot} be obtained via a deformation. Such a coproduct is able 
to produce elements in the kernel of $m$. In general, every transposition
in $S_n$ will allow to produce such a coproduct. These coproducts form in 
general no longer Hopf algebras together with the product under consideration.
However, they are needed to construct algorithmically the kernel of the product 
map. One may consider
\begin{align}
\Delta^{-}(e_i) &= e_i\otimes \Id - \Id \otimes e_i
\end{align}
and extend it as a homomorphism, forcing a bialgebra structure
\begin{align}
\Delta^{-}(m(A\otimes B)) &= m(\Delta^{-}(A)\otimes \Delta^{-}(B)) 
\end{align}
An example reads:
\begin{align}
\Delta^{-}(e_1wedge e_2) 
&= (e_1\otimes\Id-\Id\otimes e_1)(e_2\otimes\Id-\Id\otimes e_2) \nn
&= e_1\wedge e_2\otimes\Id-e_1\otimes e_2 + e_2\otimes e_1 +\Id\otimes e_1\wedge e_2 \nn
m(\Delta^{-}(e_1\wedge e_2) &= 0 \, .
\end{align}
Of course, its easy to see that $m(\Delta^{-}(A))=0$ and hence $\Delta^{-}$ 
has values in the kernel of $m$. Considering exact sequences as
\begin{align}
0 \rightarrow \textsf{Sym}_2(V^{\otimes^2})\rightarrow V\otimes V
\rightarrow V^{\wedge^2} \rightarrow 0
\end{align}
shows then that the coproducts are involved in the construction
of Schur functors, and Schur complexes, relating symmetric and antisymmetric 
powers of $V$. SVD will help to simplify and algorithmify this construction 
as will be demonstrated elsewhere, but see \cite{akin:buchsbaum:weyman:1982a}.

\subsection{Letter-place algebras, invariant theory}

Gian-Carlo Rota developed the letter-place techniques to describe invariant
theory on super algebras \cite{grosshans:rota:stein:1987a}. The Grassmann case
treated in this work is the special case where all letters, \textit{ie} formal
variables, are negatively signed, hence anticommute. The pairing between two disjoint
alphabets of letters, called letters and places, comes up with a neutral number,
behaving like a scalar. Now, let letters be $L \cong L\otimes 1$ and places 
$P\cong 1\otimes P$, a letter-place variable is given by the evaluation $[L\mid P]=
\textsf{eval}(L\otimes P) =L(P)$. The Homomorphism axiom in this case describes
the evaluation and coevaluation of invariant theory. Therefore our above given
treatment of SVD decompositions can be extended along this lines to graded
or even braided linear algebra. The biorthogonality of a spectral decomposition
should allow for more efficiency in super algebra algorithms.

\subsection{Polar decomposition of operators}

Another place, where SVD is used in disguise is the polar decomposition of 
operators. Let $A: W\rightarrow V$, $A^* : V \rightarrow W$, consider 
$A = \sqrt{A\,A^*}\, \frac{A}{\sqrt{A\,A^*}} = \rho\, \phi$. The operator
$\rho$ is a scaling operator, while $\phi$ is a 'phase'. In fact $\rho^2$
is our $D_A$ and the inverse should be taken as generalized inverse, dropping 
the kernel. If we write $A=UD^{\frac{1}{2}}_AV^T$, we get $\rho^2=AA^*=UD_AU^T$
and $\phi= UD^{-\frac{1}{2}}_A U^T UD^{\frac{1}{2}}_AV^T = UV^T$ showing
clearly that the scaling part goes into the $\rho$ while the map $UV^T$
describes the decomposition of the two tensor spaces $W$ and $V$. This is related
to our main theorem, which shows that Hopf algebras allow to compute
$\Phi=UV^T = U\circ\Delta(U)^T =m(V)\circ V^T$ using either the coproduct
or the product map on the matrix column vectors. Looking at this decomposition
in the SVD fashion allows to generalize it to singular and indefinite settings
in a meaningful way. In fact, polar decompositions might be studied using 
branching laws too.

\subsection{Numerical applications}

In numerical and computer applications, SVD is a well established method, a short
discussion is found in \cite{ablamowicz:2002a}. For applications in image 
processing, coding theory, noise reduction, latent semantic indexing, etc
see \cite{berry:dongaara:1999a,maciejowski:1989a,strang:1998a}.

\subsection{Biorthogonalization in biophysics}

A further nice application of this seminal method is the so called 'Karhunen-Loewe' 
method, actually SVD, in chaos theory and in cerebral biology, see 
\cite{haken:1996a,kelso:etal:1992a,braeuer:2002a}.

\subsection{Manifold theory -- function valued singular values}

We have skipped in the present work the complication that the duality in the
eigenspaces of $W$ and $V$ may not be mediated by matrix transposition. We
know from projective geometry and quantum field theory that coordinatizations
can be done \textit{independently} in point space and momentum space (of hyper 
planes or copoints) \cite{fauser:stoss:2004a}. This amounts to say, that we can
pair two isomorphic but not identical Grassmann algebras $V^\wedge$ and $V^{\circ_F}$,
where $\circ_F$ is another Grassmann product having a different filtration 
($F$-grading) induced by the antisymmetric bilinear form $F^\wedge$. Such a
freedom might be used to introduce a sort of 'metric' field into the branching
scheme. As an example, one might think of morphisms which connect spaces
only up to isomorphisms. Such a morphism would read in a basis
\begin{align}
m_g &: W \rightarrow V\hskip 1truecm m_g\cong [(m_g)_I^k]
\end{align}
where the indices are raised and lowered not by $\delta_I^K$ and $\delta_i^k$
but via an arbitrary, possibly function valued, $GL(V)$ element $g_{ij}$. Note 
that $g\otimes g\cong g_{IJ}$ is needed to raise/lower indices in $W$.

Having this generalization at our disposal, one might even think to put
this as a bundle structure on a manifold, which then gives function valued 
metrics $g=g(x)$, $x$ a basepoint of the manifold. We hope to investigate 
this elsewhere.

\subsection{SVD and cohomology}

Cohomological considerations proved to be extremely useful in describing
product structures in quantum field theory. The classification of such
products and their explicite evaluation in a perturbative expansion was
achieved using $\openC$-valued cohomology 
\cite{brouder:fauser:frabetti:oeckl:2002a}. However, if one consideres 
more complicated $G$-valued cohomology rings, or even cohomology monoids, 
the situation starts to get involved. Furthermore, cohomological methods are 
tied to topological invariants, hence are coarser that metric invariants. Having
the SVD available, we can ask for metric invariants and the resulting eigenvalues
carry metric information (due to the identifivation of $V$ and $V^*$).
We await therefore, that metrical information can be dealt with in the 
SVD approch better. This nourished the hope, expressed in the introduction, that
we can unveil geometrical data of non-commutative manifolds this way. 

\noindent{\bf Acknowledgement:} I would like to thank Peter Jarvis for many 
fruitful discussions about the properties of group branching laws and symmetric 
functions, which influenced this work and a critical reading of the paper. 
Rafal Ablamowicz helped me to understand the singular value decomposition in
its Clifford algebra form. Kurt Br\"auer led my attention to SVD applications 
in cerebral biology.

\small{
\bibliography{sql}
\bibliographystyle{plain}
\def\topsep{0pt}
\def\parsep{0pt plus 5pt minus 1pt}
\def\itemsep{-0.5ex}
}
\end{document}